\documentclass[twocolumn,showpacs,pra,aps,amsmath,amssymb]{revtex4}
\usepackage{mathrsfs}
\def\Tr{\operatorname{Tr}}\def\>{\rangle}\def\<{\langle}
\newtheorem{theorem}{Theorem}
\newtheorem{lemma}{Lemma}
\newtheorem{remark}{Remark}

\def\Proof{\medskip\par\noindent{\bf Proof. }}\def\qed{$\blacksquare$\par}
\def\geq{\geqslant}\def\leq{\leqslant}\def\set#1{{\sf #1}}\def\sH{\set{H}}
\def\Span{\set{Span}}\def\Supp{\set{Supp}}
\begin{document}
\title{Minimax quantum state discrimination} \author{Giacomo Mauro
D'Ariano}\email{dariano@unipv.it} \author{Massimiliano Federico 
Sacchi}\email{msacchi@unipv.it}\affiliation{{\em QUIT Group} of the
INFM, Unit\`a di Pavia} \affiliation{Universit\`a di Pavia,
Dipartimento di Fisica ``A. Volta'', via Bassi 6, I-27100 Pavia,
Italy}
\homepage{http://www.qubit.it} \author{Jonas
Kahn}\email{jokahn@clipper.ens.fr}\affiliation{Universit\'e Paris-Sud
11, D\'epartement de Math\'ematiques, B\^at 425, 91405 Orsay Cedex, 
France}\date{\today}

\begin{abstract} 
  We derive the optimal measurement for quantum state discrimination
  without {\em a priori} probabilities, i.~e. in a minimax strategy
  instead of the usually considered Bayesian one. We consider both
  minimal-error and unambiguous discrimination problems, and provide
  the relation between the optimal measurements according to the two
  schemes. We show that there are instances in which the minimum risk
  cannot be achieved by an orthogonal measurement, and this is a
  common feature of the minimax estimation strategy.
\end{abstract}

\date{\today}
\maketitle
\section{Introduction}
Since the pioneering work of Helstrom \cite{hel} on quantum hypothesis
testing, the problem of discriminating nonorthogonal quantum states
has received much attention \cite{rev12}, with some experimental
verifications as well \cite{exper}. The most popular scenarios are the
minimal-error probability discrimination \cite{hel}, where each
measurement outcome selects one of the possible states and the error
probability is minimized, and the optimal unambiguous discrimination
of linearly independent states\cite{unam}, where unambiguity is paid
by the possibility of getting inconclusive results from the
measurement.  The problem of discrimination has been addressed also
for bipartite quantum states, with both global joint measurements and
local measurements with classical communication\cite{walg}.  The
concept of distinguishability can be applied also to all physically
allowed transformations of quantum states, and in fact, more recently,
the problem of discrimination has been considered for unitary
transformations \cite{CPR} and more general quantum operations
\cite{unp}. In all the above mentioned discrimination problems, a
Bayesian approach has always been considered, with given {\em a
priori} probability distribution for the states (or operations) to be
discriminated.
  
\par In this paper, we consider the problem of optimal discrimination
of quantum states in the minimax approach. In this strategy no prior
probabilities are given.  The relevance of this approach is both
conceptual, since for a frequentist statistician the {\em a priori}
probabilities have no meaning, and practical, because the prior
probabilities may be actually unknown, as in a noncooperative
cryptographic scenario.  We will derive the optimal measurement for
minimax state discrimination for both minimal-error and unambiguous
discrimination problems. We will also provide the relation between the
optimal measurements according to the minimax and the Bayesian
strategies.  We will show that, quite unexpectedly, there are
instances in which the minimum risk can be achieved only by non
orthogonal positive operator-valued measurement (POVM), and this is a
common feature of the minimax estimation strategy.

\par The paper is organized as follows. In Sec. II we pose the problem
of discrimination of two quantum states in the minimax scenario. Such
an approach is equivalent to a minimax problem, where one should
maximise the smallest of the two probabilities of correct detection
over all measurement schemes. For simplicity we will consider equal
weights (i.e.  equal prices of misidentifying the states), and we will
provide the optimal measurement for the minimax discrimination, along
with the connection with the optimal Bayesian solution.  As mentioned,
a striking result of this section is the existence of couples of mixed
states for which the optimal minimax measurement is unique and {\em
nonorthogonal}.  In Sec. III we generalize the results for two-state
discrimination to the case of $N \geq 2$ states and arbitrary
weights. First, we consider the simplest situation of the covariant
state discrimination problem.  Then, we address the problem in
generality, resorting to the related convex programming method. In
Sec. IV we provide the solution of the minimax discrimination problem
in the scenario of unambiguous discrimination.  The conclusions of the
paper are summarized in Sec. V.

\section{Optimal minimax discrimination of two quantum states}
We are given two states $\rho_1$ and $\rho_2$, generally mixed, and we
want to find the optimal measurement to discriminate between them in a
minimax strategy. The measurement is described by a POVM with two
outcomes, namely $\vec P \equiv (P_1, P_2)$, where $P_i $ for $i=1,2$
are nonnegative operators satisfying $P_1+P_2 =I$.  \par In the
usually considered Bayesian approach to the discrimination problem,
the states are given with {\em a priori} probability distribution
$\vec a \equiv (a_1, a_2)$, respectively, and one looks for the POVM
that minimizes the average error probability
\begin{equation}
p_E=a_1 \hbox{Tr}[\rho _1 P_2] + a_2 \hbox{Tr}[\rho_ 2 P_1]. 
\end{equation}
The solution can then be achieved by taking the orthogonal POVM made
by the projectors on the support of the positive and negative part of
the Hermitian operator $a_1 \rho_ 1 -a_2 \rho _2$, and hence one has
\cite{hel}
\begin{equation}
p^{(Bayes)}_E= \frac 12 \left (1 -\Vert a_1 \rho_ 1 -a _2 \rho _2 \Vert _1
\right ),\label{pest}
\end{equation}
where $\Vert A\Vert _1 $ denotes the trace norm of $A$.
\par In the minimax problem, one does not have {\em a priori} 
probabilities. However, one defines the error probability
$\varepsilon_i(\vec P)=\Tr[\rho_i(I-P_i)]$ of failing to identify
$\rho_i$.  The optimal minimax solution consists in finding the
POVM that achieves the minimax
\begin{equation}
\varepsilon =\min_{{\vec
P}}\max_{i=1,2}\varepsilon_i(\vec P),
\end{equation}
or equivalently, that maximizes the smallest of the probabilities of correct detection
\begin{equation}
1-\varepsilon 
=\max_{{\vec P}}\min_{i=1,2}[1-\varepsilon_i(\vec P)] =\max_{{\vec
P}}\min_{i=1,2}\Tr[\rho_iP_i].
\end{equation}
The minimax and Bayesian strategies of discrimination are connected by
the following theorem.
\begin{theorem}\label{l:bayes}
  If there is an {\em a priori} probability $\vec a=(a_1,a_2)$ for the
  states $\rho_1$ and $\rho_2$, and a measurement ${\vec B}$ that
  achieves the optimal Bayesian average error for $\vec a$, with equal
  probabilities of correct detection, i.e.
\begin{equation}
\Tr[\rho_1 B_1]=\Tr[\rho_2 B_2],\label{equalB}
\end{equation}
then $\vec B$ is also the solution of the minimax discrimination problem.
\end{theorem}
\Proof In fact, suppose on the contrary that there exists a POVM $\vec
P$ such that $\min_{i=1,2}\Tr[\rho_iP_i] > \min_{i=1,2}\Tr[\rho_iB_i]$.
Due to assumption (\ref{equalB}) one has $\Tr[\rho_i P_i] >\Tr[\rho_i
B_i]$ for both $i=1,2$, whence
\begin{equation}
\sum_i a_i\Tr(\rho_iP_i) >\sum_i a_i\Tr(\rho_iB_i)
\end{equation}
which contradicts the fact that $\vec B$ is optimal for $\vec
a$.\qed 
\par The existence of an optimal $\vec B$ as in Theorem \ref{l:bayes}
will be shown in the following. 
\par First, by labeling with $\vec P^{(a)}$ an optimal POVM for the
Bayesian problem with prior probability distribution $\vec a=(a,1-a)$,
and defining
\begin{equation}
\chi(a,\vec P)\doteq a\Tr(\rho_1P_1) +(1-a)\Tr(\rho_2P_2),
\end{equation}
we have the following lemma.  
\begin{lemma} 
The function $f(a)\doteq\Tr(\rho_1P_1^{(a)})-\Tr(\rho_2P_2^{(a)})$ is
monotonically nondecreasing, with minimum value $f(0)\leq 0$, and maximum
value $f(1)\geq 0$.
\end{lemma} 
In fact, consider $\vec P^{(a)}$ and $\vec P^{(b)}$ for two values $a$
and $b$ with $a<b$ and define $\vec D=\vec P^{(b)}-\vec P^{(a)}$. Then
\begin{equation}
\begin{split}
\chi(a,\vec P^{(b)})&=\chi(a,\vec P^{(a)})+\chi(a,\vec D)\\
\chi(b,\vec P^{(a)})&=\chi(b,\vec P^{(b)})-\chi(b,\vec D).
\end{split}
\end{equation}
Now, since $\chi(a,\vec P^{(a)})$ is the optimal probability of
correct detection for prior $a$, and analogously $\chi(b,\vec
P^{(b)})$ for prior $b$, then $\chi(a,\vec D)\leq 0$ and $\chi(b,\vec
D)\geq 0$, and hence
\begin{equation}
0\le\chi(b,\vec D)-\chi(a,\vec D)=(b-a)[\Tr(\rho_1D_1)-\Tr(\rho_2D_2)].
\nonumber 
\end{equation}
It follows that  $\Tr(\rho_1D_1)\geq \Tr(\rho_2D_2)$, namely 
\begin{equation}
\Tr(\rho_1P_1^{(b)})-\Tr(\rho_1P_1^{(a)})\geq 
\Tr(\rho_2P_2^{(b)})-\Tr(\rho_2P_2^{(a)})
\end{equation}
or, equivalently,
\begin{equation}
\Tr(\rho_1P_1^{(b)})-\Tr(\rho_2P_2^{(b)})\geq \Tr(\rho_1P_1^{(a)})-\Tr(\rho_2P_2^{(a)}).
\label{equiv}
\end{equation}
Equation (\ref{equiv}) states that the function $f(a)$ is
monotonically nondecreasing. Moreover, for $a=0$ the POVM detects only
the state $\rho_2$, whence $\Tr(\rho_2P_2^{(0)})=1$, and one has
$f(0)=-1 + \Tr [\rho _1 P_1^{(0)}]\leq 0$.  Similarly one can see that
$f(1)\geq 0$.\qed 

We can now prove the following theorem.
\begin{theorem}\label{l:existsB}
  An optimal $\vec B$ as in Theorem \ref{l:bayes} always exists.
\end{theorem}
\Proof Consider the value $a_0$ of $a$ where $f(a)$ changes its sign
from negative to positive, and there take the left and right limits
\begin{equation}
\vec P^{(\mp)}=\lim_{a\to a_0^\mp}\vec P^{(a)}.
\end{equation}
For $f(a_0^+)=f(a_0^-)=0$ just define $\vec B=\vec P^{(a_0)}$. For
$f(a_0^+)>f(a_0^-)$ define the POVM $\vec B$
\begin{equation}
\vec B=\frac{f(a_0^+)\vec P^{(-)}-f(a_0^-)\vec P^{(+)}}{f(a_0^+)-f(a_0^-)}.
\end{equation}
In fact, one has 
\begin{equation}
\begin{split}
& \Tr [\rho _1 B_1] -\Tr [\rho _2 B_2]= [f(a_0^+)-f(a_0^-)]^{-1}
  \times 
\\&
  \{\Tr [\rho _1 P_1^{(-)} -\rho _2
    P_2^{(-)}]f(a_0^+) - \\&
\Tr [\rho _1 P_1^{(+)} -\rho _2
    P_2^{(+)}]f(a_0^-)\}=0\,,
\end{split}
\end{equation}
namely Eq. (\ref{equalB}) holds. 
\qed

Notice that the value $a_0$ is generally not unique, since the
function $f(a)$ can be locally constant.  However, on the Hilbert
space $\Supp(\rho _1) \cup \Supp (\rho _2) $, the optimal POVM for the
minimax problem is unique, apart from the very degenerate case in
which $D=a_0\rho_1 - (1-a_0)\rho_2$ has at least two-dimensional
kernel.  In fact, upon denoting by $\Pi _+$ and $K$ the projector on
the strictly positive part and the kernel of $D$, respectively, any
Bayes optimal POVM is written $(B_1=\Pi_+ +K',\ B_2=I-B_1 )$, with
$K'\leq K$. Since for the optimal minimax POVM we need $\Tr [\rho _1
B_1]=\Tr [\rho _2 B_2]$, one obtains $\Tr[(\rho_1
+\rho_2)K']=1-\Tr[(\rho_1+\rho _2)\Pi_+]$, which has a unique solution
$K'= \alpha K$ if $K$ is a one-dimensional projector.

\begin{remark} 
For two pure states the optimal POVM for the minimax discrimination is
  orthogonal and unique (up to trivial completion of $\Span \{|\psi
  _i\>\}_{ i=1,2 }$ to the full Hilbert space of the quantum system).
\end{remark}
In fact, on the space $\Span \{|\psi _i\>\}_{ i=1,2 }$ the optimal
  Bayes measurement is always orthogonal and unique for any prior
  probability, hence there exists an optimal POVM for the minimax
  discrimination that coincides with the optimal Bayesian one, which
  is orthogonal. Uniqueness of the minimax optimal POVM follows from
  the considerations after the proof of Theorem \ref{l:existsB} when
  restricting to the subspace spanned by the two states.
\begin{remark}\label{r:nonortho} 
There are couples of mixed states for which the optimal minimax POVM
is unique and nonorthogonal.
\end{remark}
For example, consider the following states in dimension two
\begin{equation}
\rho_1=\begin{bmatrix}1 & 0\\ 0 & 0\end{bmatrix},\quad
\rho_2=\begin{bmatrix}\frac{1}{2} & 0\\ 0 & \frac{1}{2}\end{bmatrix}.
\end{equation}
Then an optimal minimax POVM is given by
\begin{equation}\label{nonorthofreq}
P_1=\begin{bmatrix}\frac{2}{3} & 0\\ 0 & 0\end{bmatrix},\quad
P_2=\begin{bmatrix}\frac{1}{3} & 0\\ 0 & 1\end{bmatrix}.
\end{equation}
In fact, clearly there is an optimal POVM of the diagonal form. We
need to maximize $\min_{i=1,2}\Tr[\rho_iP_i]$, whence, according to
Theorem \ref{l:existsB}, we need to maximize $\Tr[\rho_1P_1]$ with the
constraints $\Tr[\rho_1P_1]=\Tr[\rho_2P_2]$ and $P_2=I-P_1$.  Such an
optimal POVM is unique, otherwise there would exists a convex
combination $a_0\rho_1-(1-a_0)\rho_2$ with kernel at least
two-dimensional, which is impossible in the present example (see
comments after the proof of Theorem \ref{l:existsB}).

\par Notice that when the optimal POVM for the minimax strategy is
unique and nonorthogonal, then there is a prior probability
distribution $\vec a$ for which the optimal POVM for the Bayes problem
is not unique, and the nonorthogonal POVM that optimizes the minimax
problem is also optimal for the Bayes' one. In the example of Remark
\ref{r:nonortho} the optimal POVM (\ref{nonorthofreq}) is also optimal
for the Bayes problem with $\vec a=(\frac{1}{3} ,\frac{2}{3})$ as one
can easily check.  However, in the Bayes case one can always choose an
optimal orthogonal POVM, whereas in the minimax case you may have to
choose a non-orthogonal POVM.

\par Finally, notice that, unlike in the Bayesian case, the optimal
POVM for the minimax strategy may also be not extremal.

\section{Optimal minimax discrimination of $N \geq 2$ quantum states}
We now consider the easiest case of discrimination with more than two
states, namely the discrimination among a covariant set. In a fully
covariant state discrimination, one has a set of states $\{\rho_i\}$
with $\rho_i=U_i\rho_0U_i^\dag$ $\forall i$, for fixed $\rho_0$ and
$\{U_i\}$ a (projective) unitary representation of a group.  In the
Bayesian case full covariance requires that the prior probability
distribution $\{a_i\}$ is uniform. Then, one can easily prove (see,
for example, Ref. \cite{Holevobook}) that also the optimal POVM is
covariant, namely it is of the form $P_i=U_iKU_i^\dag$, for suitable
fixed operator $K\geq 0 $.

\begin{theorem}\label{cov} 
For a fully covariant state discrimination problem, there is an
optimal measurement for the minimax strategy that is covariant,
and coincides with an optimal Bayesian measurement.
\end{theorem}
\Proof A covariant POVM $\{P_i\}$ gives a probability
$p=\Tr[\rho_iP_i]$ independent of $i$.  Moreover, there always exists
an optimal Bayesian POVM that is covariant and maximizes $p$, which
then is also the maximum over all POVM's of the average probability of
correct estimation $\overline{\Tr[\rho_iP_i]}$ for uniform prior
distribution \cite{Holevobook}. Now, suppose by contradiction that
there exists an optimal minimax POVM $\{P'_i\}$ maximizing
$p'=\min_i\Tr[\rho_iP'_i]$, for which $p'>p$. Then, one has
$p<p'\leq\overline{\Tr[\rho_iP_i']}$, contradicting the assertion that
an optimal Bayesian POVM maximizes $\overline{\Tr[\rho_iP_i]}$ over
all POVM's. Therefore, $p=p'$, and the covariant Bayesian POVM also
solves the minimax problem.\qed

Notice that in the covariant case also for any optimal minimax POVM
$\{P_i\}$ one has $\Tr[\rho_iP_i]$ independent of $i$, since the
average probability of correct estimation is equal to the minimum one.

\par In the following we generalize Theorem \ref{l:bayes} for two
states to the case of $N \geq 2 $ states and arbitrary weights. We
have
\begin{theorem}
  For any set of states $\{\rho _i \}_{2 \leq i \leq N}$ and any set
  of weights $w_{ij}$ (price of misidentifying $i$ with $j$) the
  solution of the minimax problem
\begin{equation}
r=\inf _{\vec P}\sup _i \sum_j w_{ij}\Tr [\rho _i P_j]\;\label{11}
\end{equation}
is equivalent to the solution of the problem 
\begin{equation}
r=\max _ {\vec a} r_B(a),\label{12}
\end{equation}
where $r_B(\vec a)$ is the Bayesian risk
\begin{equation}
r_B(\vec a) \doteq\max _{\vec P}
\sum _i a_i \sum_j w_{ij}\Tr 
[\rho _i P_j].\label{12b}
\end{equation}
\end{theorem}
\Proof The minimax problem in Eq. (\ref{11}) is equivalent to look for
the minimum of the real function $\delta=f(\vec P)$ over $\vec P$,
with the constraints
\begin{eqnarray}
&\sum _j w_{ij} \Tr[\rho _i P_j]\leq \delta,\quad&\forall i \nonumber \\&  
P_j\geq 0,\quad&\forall j \nonumber \\ & 
\sum _j P_j =I.&\label{constraints}
\end{eqnarray}
Upon introducing the Lagrange multipliers: 
\begin{equation}
\begin{split}
\mu_i\in{\mathbb R}^+\,,\quad&\forall i\\
0\le Z_i\in M_d({\mathbb C}),\quad&\forall i\\
Y^\dag=Y\in M_d({\mathbb C}),&\label{set}
\end{split}
\end{equation}
$M_d({\mathbb C})$ denoting the $d\times d$ matrices on the complex
field, the problem is equivalent to
\begin{eqnarray}
&&r=\inf _{\vec P, \delta }{\sup_{\vec \mu, \vec Z, Y}}\!\!'  \ 
l(\vec P, \delta , \vec \mu , \vec Z ,Y),
\nonumber \\& & l(\vec P, \delta , \vec \mu , \vec Z ,Y)\doteq 
\delta +\sum _i [\mu _i (\sum _j w_{ij} \Tr [\rho _i P_j]- \delta )] 
\nonumber \\& & -
\sum _i \Tr [Z_i P_i] + \Tr[Y(I -\sum _i P_i) ],\label{conv}
\end{eqnarray}
where $\sup '$ denotes the supremum over the set defined in Eqs.
(\ref{set}). The problem is convex [namely both the function $\delta $
and the constraints (\ref{constraints}) are convex] and meets Slater's
conditions \cite{Vandenberghe} (namely one can find values of $\vec P$
and $\delta $ such that the constraints are satisfied with strict
inequalities), and hence in Eq. (\ref{conv}) one has
\begin{equation}
\inf _{\vec P, \delta }{\sup _{\vec \mu, \vec Z, Y}}\!\!' 
\ l(\vec P, \delta , \vec \mu , \vec Z ,Y)
=
{\max _{\vec \mu, \vec Z, Y}}' 
\inf _{\vec P, \delta }\ l(\vec P, \delta , \vec \mu , \vec Z ,Y).\label{inv}
\end{equation}
It follows that 
\begin{equation}
r={\max _{\vec \mu, \vec Z, Y}}' \Tr Y
\end{equation}
under the additional constraints
\begin{eqnarray}
&&\sum _i \mu _i =1 \,,\nonumber \\& & 
\sum _i w_{ij}\mu _i\rho _i -Z_j -Y =0\;, \qquad \forall j.
\end{eqnarray}
The constraints can be rewritten as 
\begin{eqnarray}
&&\mu _i \geq 0 \,,\qquad \sum _i \mu _i =1 \,,\nonumber \\& & 
Y \leq \sum _i w_{ij}\mu _i\rho _i\;, \qquad \forall j.
\end{eqnarray}
Now, notice that for the Bayesian problem with prior $\vec a$, along
the same reasoning, one writes the equivalent problem
\begin{equation}
r_B(\vec a) ={\max _Y}' \Tr Y,
\end{equation}
with the constraint 
\begin{eqnarray}
&&\sum _i w_{ij}a_i\rho _i -Z_j -Y =0\;, \qquad \forall j\\
&&a_i \geq 0 \,,\qquad \sum _i a_i =1 \,,\nonumber \\& & 
Y \leq \sum _i w_{ij}a_i\rho _i\;, \qquad \forall j,
\end{eqnarray}
which is the same as the minimax problem, with the role of the
Lagrange multipliers $\{\mu_i\}$ now played by the prior probability
distribution $\{a_i\}$.\qed Clearly, a POVM that attains $r$ in the
minimax problem (\ref{11}) actually exists, being the infimum over a
(weakly) compact set---the POVM convex set---of the (weakly)
continuous function $\sup_i\sum_j w_{ij}\Tr [\rho _i P_j]$.

\section{Optimal minimax unambiguous discrimination}

In this section we consider the so-called unambiguous discrimination
of states \cite{unam}, namely with no error, but possibly with an
inconclusive outcome of the measurement. We focus attention on a set
of $N$ pure states $\{\psi_i\}_{i\in \set{S}}$. In such a case, it is
possible to have unambiguous discrimination only if the states of the
set $\set{S}$ are linearly independent, whence there exists a
biorthogonal set of vectors $\{|\omega_i\>\}_{i\in \set{S}}$, with
$\<\omega_i|\psi_j\>=\delta_{ij}$, $\forall i,j\in\set{S}$.  We will
conveniently restrict our attention to $\Span\{|\psi _i\>\}_{ i\in
\set{S}}\equiv\sH $ (otherwise one can trivially complete the optimal
POVM for the subspace as a POVM for the full Hilbert space of the
quantum system). While in the Bayes problem the probability of
inconclusive outcome is minimized, in the minimax unambiguous
discrimination we need to maximize $\min_i\<\psi_i|P_i|\psi_i\>$ over
the set of POVM's with $\<\psi_i|P_j|\psi_i\>=0$ for $i\neq j \in
\set{S}$, and the POVM element that pertains to the inconclusive
outcome will be given by $P_{N+1}=I -\sum_{i\in \set{S}}P_i$.  We have
the following theorem.
\begin{theorem}\label{t:unamb} The optimal minimax unambiguous 
discrimination of $N$ pure states $\{\psi_i\}_{i\in \set{S}}$ is
  achieved by the POVM
\begin{equation}\label{unamfr1}
\begin{split}
P_i=&\kappa|\omega_i\>\<\omega_i|,\qquad i\in\set{S}\,,\\
P_{N+1}=&I-\sum_{i\in \set{S}}P_i\,,
\end{split}
\end{equation}
where $\kappa$ is given by
\begin{equation}\label{unamfr2}
\kappa^{-1}=\text{\em max eigenvalue of
}\sum_{i\in\set{S}}|\omega_i\>\<\omega_i|\,.
\end{equation}
\end{theorem}
\Proof We need to maximize $\min_i\<\psi_i|P_i|\psi_i\>$ over the set
of POVM's with $\<\psi_i|P_j|\psi_i\>=0$ for $i\neq j \in \set{S}$,
whence clearly $P_j=\kappa_j|\omega_j\>\<\omega_j|$. Then the problem
is to maximize $\min_{i\in\set{S}}\kappa_i$. This can be obtained by
taking $\kappa_i=\kappa$ independent of $i$ and then maximizing
$\kappa$. In fact, if there is a $\kappa_i>\kappa_j$ for some $i,j$,
then we can replace $\kappa_i$ with $\kappa_j$, and iteratively we get
$\kappa_i=\kappa$ independently of $i$.  Finally, the maximum $\kappa$
giving $P_{N+1}\ge 0$ is the one given in the statement of the
theorem. \qed

As regards the unicity of the optimal POVM, we can show the following.  
\begin{theorem}\label{t:unamb2}  The optimal POVM of Theorem \ref{t:unamb} 
is non-unique if and only if $|\omega_i\>\in\Supp(P_{N+1})$ for
  some $i\in\set{S}$.
\end{theorem}
\Proof In fact, if there exists an $i\in\set{S}$ such that
$|\omega_i\>\in\Supp(P_{N+1})$, this means that there exists
$\varepsilon>0$ such that $\varepsilon|\omega_i\>\<\omega_i|\le
P_{N+1}$.  Then the following is a POVM
\begin{equation}
\begin{split}
Q_j&=P_j,\quad\mbox{for }j\neq i\\
Q_i&=P_i+\varepsilon|\omega_i\>\<\omega_i|,\\
Q_{N+1}&=P_{N+1}-\varepsilon|\omega_i\>\<\omega_i|,\\
\end{split}
\end{equation}
and is optimal as well. Conversely, if there exists another
equivalently optimal POVM $\{Q_j\}$, then there exists an
$i\in\set{S}$ such that $Q_i>P_i$ (since both are proportional to
$|\omega_i\>\<\omega_i|$, and $\min_i\<\psi_i|Q_i|\psi_i\>$ has to be
maximized). Then $|\omega_i\>\in\Supp(P_{N+1})$.\qed 

%\medskip 

\par
When the optimal POVM according to Theorem \ref{t:unamb2} is not
unique, one can refine the optimality criterion in the following way.
Define the set $\set{S}_1\subset \set{S}$ for which one has
$|\omega_i\>\in\Supp(P_{N+1})$. Denote by $\mathfrak{P}_1$ the set of
POVM's that are equivalently optimal to those of Theorem
\ref{t:unamb}. Then define the set of POVM's $\mathfrak{P}_2\subset
\mathfrak{P}_1$ that maximizes $\min_{i\in \set{S} _1}\< \omega
_i|P_i|\omega_i\>$. In this way one iteratively reach a unique optimal
POVM, which is just the one given in Eqs. (\ref{unamfr1}) and
(\ref{unamfr2}).

\section{Conclusions}
In conclusion, we have considered the problem of optimal
discrimination of quantum states in the minimax strategy. This
corresponds to maximising the smallest of the probabilities of correct
detection over all measurement schemes. We have derived the optimal
measurement both in the minimal-error and in the unambiguous
discrimination problem for any number of quantum states.  The relation
between the optimal measurement and the optimal Bayesian solutions has
been given.  Differently from the Bayesian scenario, we have shown
that there are instances in which the minimum risk cannot be achieved
by an orthogonal measurement. Finally, in the unambiguous
discrimination problem, we have shown a refinement of the minimax
problem that leads always to a unique optimal minimax measurement.

\section*{Acknowledgments}
We thank G. Chiribella for correcting the original proof of Theorem
\ref{cov}.  Support from INFM through the project PRA-2002-CLON, and
from EC and MIUR through the cosponsored ATESIT project IST-2000-29681
and Cofinanziamento 2003 is acknowledged.

\end{document}